\UseRawInputEncoding
\documentclass[%
 preprint, superscriptaddress,
 amsmath,amssymb,
 aps, physrev,
]{revtex4-2}

\usepackage{graphicx}% Include figure files
\usepackage{dcolumn}% Align table columns on decimal point
\usepackage{bm}% bold math
\usepackage{multirow}
\usepackage{colortbl}
\usepackage[dvipsnames]{xcolor}
\begin{document}

\title{\textbf{The effects of temperature and viscosity on the metachronal swimming of crustaceans. }}

\author{Adrian Herrera-Amaya}
\author{Nils B. Tack}
\affiliation{Center for Fluid Mechanics, School of Engineering, Brown University, Providence, RI, 02912, USA}
\author{Zhipeng Lou}
\author{Chengyu Li}
\affiliation{Department of Mechanical and Aerospace Engineering, Case Western Reserve University, Cleveland, OH, 44106, USA}
\author{Monica M. Wilhelmus}
\email{Contact author: mmwilhelmus@brown.edu}
\affiliation{Center for Fluid Mechanics, School of Engineering, Brown University, Providence, RI, 02912, USA}

\date{\today}% It is always \today, today,
             %  but any date may be explicitly specified

\begin{abstract}
Temperature changes as small as $3 ^\circ$C have been observed to significantly impact how self-propelled organisms move through their environment, especially for those inhabiting the transitional flow regime in which both viscous and inertial effects are important. Nonetheless, many oceanic species can successfully migrate across temperature changes in the order of $20 ^\circ $C, corresponding to $40 \%$ differences in viscosity, via metachronal propulsion, suggesting that this propulsion mechanism is resilient to drastic changes in water column properties. We investigate marsh grass shrimp (\textit{Palaemon vulgaris}) as a model organism to explore the combined physical and physiological effects on their locomotion at natural seasonal temperature extremes ($6^\circ - 20^\circ$C). Experimentally, we manipulate temperature and viscosity independently to isolate physical and physiological effects. We then use the shrimp morphology and gait data to inform a computational fluid dynamics parametric study to estimate the force-to-power ratios of varying viscosity and beat frequencies through naturally occurring extremes. Our research demonstrates that shrimp do not modify their gait parameters to naturally occurring viscosity changes, and their swimming performance is impacted by less than $9 \% $. The robustness of the metachronal gait is evidence of the ecological success of shrimp-like organisms in all climates, from the tropics to pole waters and inland freshwater.
\end{abstract}

\maketitle
\newpage
%\tableofcontents

\section{\label{sec:level1}Introduction}

Temperatures in the oceans can vary up to $30^\circ$C, due to a variety of factors like local weather, seasonal anomalies, and regional variability (e.g., tropic, polar, surface, and deep waters) \cite{abraham_review_2013}. These changes have been observed to strongly influence animal swimming performance on physiological and physical levels \cite{podolsky_separating_1993}. On a physiological level, temperature mainly affects muscle biochemical reactions, leading to slower motion at low temperatures \cite{koumoundouros_effect_2002}. Meanwhile, temperature also changes the physical properties of water (i.e., viscosity), which alters fluid-structure interactions \cite{larsen_effect_2008}. In this sense, underwater propulsion is a highly coupled problem between physiological and physical effects, in which animals have developed different locomotion adaptations.
 
The coupled physiological and physical effect of temperature has long been the focus of scientific studies, albeit solely within the extremes of low or high Reynolds numbers ($Re$). At low $Re$ $(< 1)$, viscous forces dominate; at high $Re$, inertial forces are more important. In both regimes, there are examples of swimmers that modify their swimming gait in response to environmental conditions \cite{gemmell_compensatory_2013,fuiman_what_1997} and cases in which there are no adaptations \cite{johnson_partitioning_1998}. In this study, we focus on the impact of temperature on metachronal locomotion --- a widespread swimming mode among many of the most abundant aquatic organisms \cite{byron_metachronal_2021}. For instance, only the marine subphylum crustacea has more than 30,000 different species \cite{noauthor_wonders_nodate}. In contrast to high or low $Re$ cases, when swimming at intermediate Reynolds numbers on the order of 1-1000, inertia and viscosity both play a key role in propulsion. Given the exceptional functionality and widespread presence of metachronal swimmers in the global ocean, we investigate how these organisms respond to changing environmental conditions. 

Among metachronal marine species, shrimp-like organisms (i.e., caridoid facies crustaceans) \cite{ruszczyk_trends_2022} stand out for their maneuverability \cite{murphy_metachronal_2011}, swimming endurance \cite{bianchi_diel_2013}, and their widespread distribution around the globe \cite{meland_taxonomic_2015}. Such capabilities have resulted in multiple studies on their propulsion dynamics using animal experiments \cite{murphy_hydrodynamics_2013,hanson_mantis_2023,garayev_metachronal_2021}, mathematical models \cite{alben_coordination_2010,granzier-nakajima_numerical_2020}, and robotic prototypes \cite{ford_hydrodynamics_2019,oliveira_santos_pleobot_2023,tack_going_2024}. However, their swimming performance in response to environmental conditions has yet to be explored. Shrimp-like organisms are present in a diverse range of climates\cite{meland_taxonomic_2015,mantovani_contrasting_2021}, and some species like krill are exposed to significant daily temperature changes when performing diel vertical migrations spanning hundreds of meters \cite{bianchi_diel_2013}; suggesting that their locomotion is either resilient or adaptable to the accompanying physical and physiological effects. 

In this study, we chose marsh grass shrimp (\textit{P. vulgaris}) as a model organism to study the temperature effects on the forward swimming gait of shrimp-like organisms. Previous studies on the temperature-dependent swimming of aquatic invertebrates have focused on comparing locomotion kinematics. Still, there has been no integrative study (i.e., kinematics and forces) directly linking the viscosity effect to locomotion. For the first time, we establish the role of temperature-driven viscosity change on locomotion through a combination of high-speed videography and computational fluid dynamics. We separated the physiological and physical effects of temperature by artificially altering seawater viscosity at naturally occurring temperature extremes $(6^\circ - 20^\circ$C). The primary aim of our study is to investigate the compensatory mechanisms shrimps use to maintain adequate performance across varying environmental conditions. By integrating morphology and gait kinematic data into a three-dimensional Computational Fluid Dynamics model (CFD), we aim to simulate steady-state forward swimming under conditions of varying fluid viscosity and appendage beat frequencies. This approach will help us understand the impact of changes in water properties, such as viscosity, on shrimp propulsion performance. Additionally, our study examines how shrimps respond to water temperature changes, focusing on physiological adjustments rather than dynamic alterations to their swimming gait. The findings from this research could provide significant insights into the resilience of metachronal crustaceans, particularly in the context of global climate change\cite{meland_taxonomic_2015}, and highlight their ability to thrive in diverse climates worldwide.

\section{Methods}

\subsection{Experimental approach}

Marsh grass shrimp (\textit{Palaemonetes vulgaris}) (n = 21; body length $L_b$ = 3.11 $\pm$ 0.25 cm) were collected in June 2022 from Narragansett Bay (Rocky Point State Park, Warwick, RI, USA). The shrimp were divided into three groups of 7 animals and were acclimated and maintained in three separate 38-liter aerated aquaria with a salinity of 30 ppt. One tank was kept at $6^\circ$C (high viscosity, low temperature, T06) and the remaining two tanks at $20^\circ$C (low viscosity, high temperature, T20). These temperatures were chosen since they corresponded to the known temperature extremes this species tolerates in their native environment \cite{nixon_one_2004}. Capture and experiments were conducted in accordance with the laws of the State of Rhode Island.

The physical and metabolic effects during steady swimming were investigated through the independent manipulation of water temperature and viscosity using three treatments: (i) at $6^\circ$C (low temperature, high viscosity), (ii) $T = 20^\circ$C (high temperature, low viscosity), and (iii) at $20^\circ$C with addition of Polyvinyl pyrrolidone (PVP) to match the viscosity at $6^\circ$C (high temperature, high viscosity). Viscosity was adjusted by dissolving 30 ppt artificial seawater. A routine viscometer was used to determine the concentration of PVP solution that matched the viscosity of $20^\circ$C seawater ($\nu = 1.017$ mm $^{2}$ s$^{-1}$) to that of $6^\circ$C seawater ($\nu = 1.519$ mm $^{2}$ s$^{-1}$). PVP is a suitable agent for manipulating viscosity because PVP solutions show constant viscosity over a wide range of shearing stresses \cite{baba_quantitative_1970}, are non-toxic, and have no effects on the metabolic rates of organisms \cite{podolsky_separating_1993}. To acquire morphological and swimming parameters, we recorded from the lateral view (corresponding to the left side of the animals) using a high-speed digital video camera (Fastcam Nova R2, Photron, Tokyo, Japan) at 2000 frames s$^{-1}$ and resolution of 2048 $\times$ 1472 pixels. Pleopod kinematics data were extracted from image sequences using the DLTdv8 Matlab package and a custom program in MATLAB. The $\alpha$ angle, defining the pleopod beat amplitude, was calculated by measuring the angle between the protopod and the line passing through the proximal joint of the protopods of P1 and P5 (aligned in the direction of swimming). The $\beta$ angle was measured between the protopodite and the line passing through the proximal and distal sections of the ramal structures. The beat frequency, $f$, temporal asymmetry, $Ta$, and phase lag $PL$ are all calculated from the $\alpha$ angle kinematics. By comparing the protopods angular velocity and $\alpha$ position, we can detect the beginning and end of the power and recovery strokes. To estimate the $Q_{10}$ value we used the following equation $Q_{10} = (R_2/R_1)^{(10/(T_2-T_1))}$, where $T$ is the temperature in $^\circ$C, and $R$ is the respective beat frequency.

\subsection{Numerical approach}

Due to the significance of both viscous and inertial forces in the propulsion of marsh grass shrimp, we utilize an immersed-boundary-method(IBM)-based in-house CFD solver to numerically solve the 3D viscous incompressible Navier-Stokes equations. The equations in their nondimensional form are expressed as follows:

\begin{equation}
    \frac{\partial u_i}{\partial x_i} = 0; \quad
    \frac{\partial u_i}{\partial t} + \frac{\partial (u_i u_j)}{\partial x_j} = -\frac{\partial p}{\partial x_i} + \frac{1}{\text{Re}} \frac{\partial}{\partial x_j} \left(\frac{\partial u_i}{\partial x_j}\right)
\end{equation}\\
Where $u_i$ are the velocity components, $p$ is the pressure, and $\text{Re}$ is the Reynolds number.

The equations outlined above are discretized using a cell-centered, collocated arrangement of the primitive variables on a Cartesian grid. They are solved using a finite difference-based immersed-boundary method \cite{mittal_versatile_2008}. Time integration is performed using the fractional step method, while a second-order central difference scheme is used for spatial discretization. The Eulerian form of the Navier-Stokes equations is discretized on a Cartesian mesh, and boundary conditions on the immersed boundary are applied through a ghost-cell procedure. This IBM approach, compared to boundary-conforming methods such as curvilinear grids\cite{visbal_large-eddy_2002} and finite element methods \cite{tezduyar_finite_2004}, eliminates the need for complex re-meshing algorithms, thereby significantly reducing the computational cost associated with simulating flow around complex moving boundaries. Immersed boundary methods fall into two broad categories: the continuous forcing approach and the discrete forcing approach. Our study utilizes a multi-dimensional "ghost-cell" methodology to enforce boundary conditions on the immersed boundary, which is categorized as a discrete forcing approach where the forcing is directly incorporated into the discretized Navier-Stokes equations. The motion of the immersed boundaries (shrimp pleopods) is prescribed based on image-based reconstructions. This methodology has been successfully applied in simulations of bio-inspired propulsion systems\cite{li_balance_2018,lou_wingantenna_2024,lionetti_benefits_2025}. Validations of the current in-house CFD solver are documented in our previous studies \cite{lionetti_new_2023,li_computational_2017,lionetti_aerodynamic_2022}. {Details on the shrimp numerical model can be found in \textit{Lou et al.} \cite{zhipeng_l_edge_nodate}.

Based on our simulation data, we evaluated swimming performance by integrating the forces along the pleopods in the thrust $F_T$ and lift $F_L$ directions. In this investigation, the effect of each force is presented in terms of non-dimensional coefficients, which are computed as $C_L = F_L/[0.5 \rho (U_{tip})^2 S]$ and $C_T = F_T/[0.5 \rho (U_{tip})^2 S]$. Here, $\rho$ is the water density, $U_{tip}$ is the average pleopod velocity, and S denotes the average area of the pleopod surface. The instantaneous hydrodynamic power $(P_{hydro} = \oint -(\boldsymbol{\sigma} \cdot \hat{n}) \cdot \vec{V} \,ds)$ is the rate of work done by the pleopod model, where $\oint$ denotes the pleopod surface integration, $\boldsymbol{\sigma}$ and $\vec{V}$ represent the stress tensor and the velocity vector of the fluid adjacent to the model surface and $\hat{n}$ is the vector normal to the surface. In a non-dimensional formulation the power coefficient is calculated as $C_{PW} = P_{hydro}/[0.5 \rho (U_{tip})^3 S]$. 

\section{Results}

We evaluated the locomotion of our model organism during forward swimming under three different environmental conditions: (i) $T = 6^\circ $C (low temperature, high viscosity), (ii) $T = 20^\circ C$ (high temperature, low viscosity), and (iii) $T = 20^\circ C$ with the addition of polyvinyl pyrrolidone (PVP) to match the viscosity at $6^\circ C$ (high temperature, high viscosity). We will refer to the three different environmental conditions as T06 (i), T20 (ii), and T20+PVP (iii). When comparing T20 \textit{vs.} T20+PVP, we isolate the effect of viscosity; meanwhile, T06 \textit{vs.}  T20+PVP isolates the effect of temperature. To statistically compare the swimming parameters, data sets were tested for normality (by a Shapiro-Wilk test) and equal variances (by a Levene's test). Both one-way ANOVA and t-tests were used when appropriate (Table \ref{tab: Stats}). A Kruskal-Wallis test was performed only when data sets did not pass the normality or variance assumptions. Results from the normality and variance tests are listed in the Appendixes (Tables \ref{tab:Test1} and \ref{tab:Test2}). 

\subsection{Morphological and swimming parameters} 

We characterize the forward swimming gait using three morphological, five kinematic, and two swimming variables. Morphometric parameters include the body length $(L_b)$, protopodite length $(L_p)$, and the endopodite length $(L_e)$ (Figure \ref{fig:Morphology}A). Kinematic parameters include alpha $(\alpha)$, the angle between the body line and the protopodite, beta $(\beta)$, the angle between the protopodite and the biramous distal appendage (Figure \ref{fig:Morphology}B), the beat frequency $(f)$, phase-lag $(PL)$, and temporal asymmetry $(Ta)$. The temporal asymmetry is defined as $Ta = (t_r-t_p)/(t_r+t_p)$, where $t_r$ and $t_p$ are the times for recovery and power stroke. $Ta$ is zero when the recovery and power stroke have the same duration $(t_r = t_p)$ and get closer to one as $t_p < t_r$. To compare forward swimming between the treatments, we use the mean swimming speed $(V_b)$ and the advance ratio $(J = V_b/2\alpha f \textit{l})$, an efficiency metric using the ratio of the forward speed of the animal to the mean velocity of the tip of one of the appendages. 

Tracking the center of mass during forward swimming, we observed low-temperature fluid yielding lower swimming speeds than its counterpart at higher temperature with the same viscosity (T06 \textit{vs.} T20+PVP t-test, $p = 0.01, t = -2.98, df = 12$, Figure \ref{fig: BodyKine}A).  The temperature coefficient $Q_{10}$, commonly used in biology to describe the metabolic rate of change as a consequence of increasing the temperature by $10 ^\circ$ C \cite{schmidt-nielsen_animal_1997}, is 1.52 for the mean swimming velocity--suggesting a strong temperature dependence. When comparing the viscosity effect (T20 \textit{vs.} T20+PVP), the swimming speed shows no statistical difference (t-test, $p = 0.68, t = 3.43, df = 12$). To explore the physiological impact on swimming performance, we can estimate swimming efficiency by calculating the advance ratio, $J$. In this case, we consider the kinematics of each appendage to evaluate $J$ (Figure \ref{fig: BodyKine}B). From one-way ANOVAs, we found that there were no significant differences between pleopods (T06-ANOVA, $p = <0.001, F = 10.76, df = 4$ / T20-ANOVA, $p = <0.001, F = 7.36, df = 4$ / T20+PVP-ANOVA, $p = <0.001, F = 13.19, df = 4$). Animals with a single pair of appendages have advance ratios between 0.1 and 0.6 \cite{walker_functional_2002}, for shrimp $J \approx 1$ for all the pleopods, highlighting the benefit of metachronal motion. When comparing the three treatments, we followed the logic of a two-way ANOVA. If two or more pleopods do not satisfy the null hypothesis that there is no difference in the mean ($p < 0.05$), the treatments are considered significantly different. We found no significant difference between treatments when comparing $J$ between environmental conditions (see Table \ref{tab: Stats}). The advance ratio normalizes the swimming speed by the appendage speed; having the same advance ratio tells us that shrimp maintain the relation between appendage speed and swimming speed despite environmental changes and the consequent physiological and physical effects. 

\begin{figure}[hbt!]
\centering
\includegraphics[width=\linewidth]{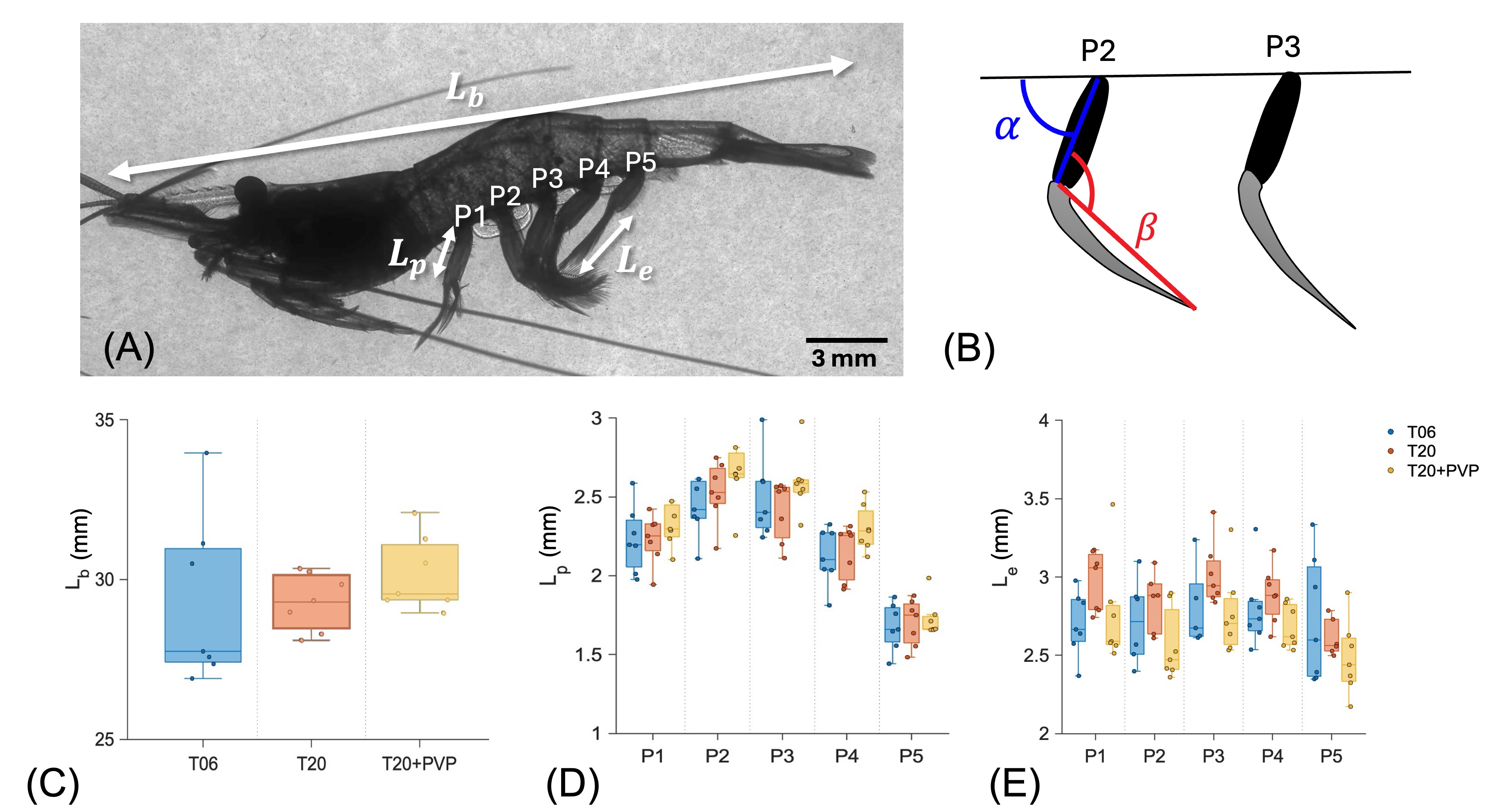}
\caption{Parameters governing the kinematics and morphology of \textit{P. vulgaris}. (A) Lateral view showing the five pleopods (labeled anteroposteriorly from P1 to P5), the body length $(L_b)$, the protopodite length $(L_p)$, and the distal biramous segment length $(L_e)$, (B) Schematic of two pleopods showing the definitions of $\alpha$, the angle between the protopodite and the body axis, and $\beta$, the angle between the protopodite and the distal biramous segment. (C) Body lengths for the animals used in the three different treatments: \textcolor{blue}{T06} $(n = 7)$, \textcolor{red}{T20} $(n = 7)$, and \textcolor{Goldenrod}{T20+PVP} $(n = 7)$. (D) Protopodite length measurements for animals in all treatments. (E) Distal biramous segment length measurements for animals in all treatments. The boxplots display the median, the lower and upper quartiles, outliers, and the minimum and maximum values that are not outliers. }
\label{fig:Morphology}
\end{figure}

\begin{figure}[hbt!]
\centering
\includegraphics[width=\linewidth]{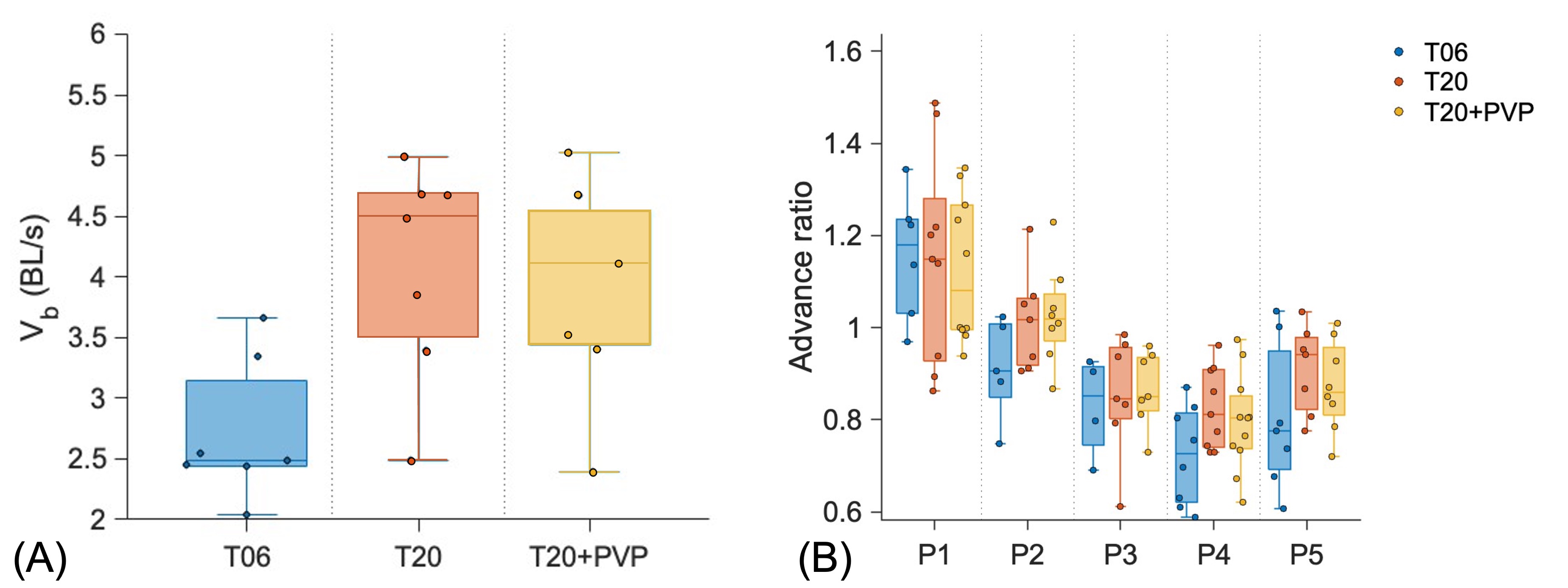}
\caption{Full body swimming kinematics. (A) Mean swimming speed $V_b$ for each recorded animal ($n = 7$). (B) Advance ratio calculation considering the kinematics of each pleopod. For each pleopod variable, each observation corresponds to a complete beat cycle. The number of observations for \textcolor{blue}{T06}: $n_{P1} = 7$, $n_{P2} = 7$, $n_{P3} = 7$, $n_{P4} = 8$, and $n_{P5} = 8$, for \textcolor{red}{T20}: $n_{P1} = 9$, $n_{P2} = 8$, $n_{P3} = 8$, $n_{P4} = 10$, and $n_{P5} = 9$, and for \textcolor{Goldenrod}{T20+PVP}: $n_{P1} = 10$, $n_{P2} = 8$, $n_{P3} = 7$, $n_{P4} = 11$, and $n_{P5} = 9$. The boxplots display the median, the lower and upper quartiles, outliers, and the minimum and maximum values that are not outliers. }
\label{fig: BodyKine}
\end{figure}

\begin{table}[hbt!]
\resizebox{\textwidth}{!}{
\begin{tabular}{|l|l|l|l|l|l|l|}
\hline
Variable             & Treatment      & P1 (p/t/df)   & P2 (p/t/df)   & P3 (p/t/df)   & P4 (p/t/df)   & P5 (p/t/df)   \\ \hline
\multirow{3}{*}{J}   & T06 vs T20     & 0.95/-0.05/13 & 0.14/1.59/10  & 0.77/0.29/9   & 0.45/2.17/15  & 0.15/1.50/12  \\ \cline{2-7} 
                     & T20 vs T20+PVP & 0.78/0.28/17  & 0.82/-0.22/13 & 0.81/-0.23/12 & 0.47/0.72/8   & 0.48/0.72/13  \\ \cline{2-7} 
                     & T06 vs T20+PVP & 0.69/0.39/14  & 0.08/-1.86/11 & 0.54/-0.62/9  & 0.16/-1.43/17 & 0.32/-1.02/13 \\ \hline
\multirow{3}{*}{\textit{f}}   & T06 vs T20     & $<$0.001/10.16/1  & 0.03/2.47/10  & $<$0.001/3.72/9   & $<$0.001/3.97/15  & \textbf{$<$0.001/8.28/1}   \\ \cline{2-7} 
                     & T20 vs T20+PVP & 0.77/0.08/1   & 0.54/-0.62/13 & 0.78/0.28/12  & 0.42/-0.81/18 & \textbf{0.86/0.03/1}   \\ \cline{2-7} 
                     & T06 vs T20+PVP & $<$0.001/10.34/1  & 0.01/-2.87/11 & $<$0.001/-3.52/9  & $<$0.001/-4.07/17 & \textbf{$<$0.001/8.37/1}   \\ \hline
\multirow{3}{*}{$\alpha$}   & T06 vs T20     & \textbf{0.23/1.39/1}  & 0.44/-0.80/10 & 0.50/0.70/9   & 0.32/-1.00/15 & 0.34/-0.98/12 \\ \cline{2-7} 
                     & T20 vs T20+PVP & \textbf{0.16/1.93/1}  & 0.56/0.58/13  & 0.54/0.61/12  & 0.02/2.52/18  & 0.50/0.69/13  \\ \cline{2-7} 
                     & T06 vs T20+PVP & \textbf{0.02/5.19/1}  & 0.14/1.54/11  & 0.98/-0.01/9  & $<$0.001/3.36/17  & 0.20/1.33/13  \\ \hline
\multirow{3}{*}{Ta}  & T06 vs T20     & 0.73/0.34/13  & 0.32/1.03/10  & \textbf{0.44/0.57/1}  & 0.35/-0.96/15 & 0.89/-0.14/12 \\ \cline{2-7} 
                     & T20 vs T20+PVP & 0.19/-1.33/17 & 0.91/-0.11/13 & \textbf{0.27/1.18/1}  & 0.77/-0.29/18 & 0.43/-0.80/13 \\ \cline{2-7} 
                     & T06 vs T20+PVP & 0.09/-1.78/14 & 0.30/-1.06/11 & \textbf{0.70/0.14/1}  & 0.49/0.69/17  & 0.29/1.08/13  \\ \hline
\multirow{3}{*}{$\beta$}   & T06 vs T20     & \textbf{$<$0.001/6.72/1}  & $<$0.001/4.41/10  & $<$0.001/3.87/9   & 0.01/2.63/15  & 0.03/2.4/12   \\ \cline{2-7} 
                     & T20 vs T20+PVP & \textbf{$<$0.001/8.17/1}  & 0.14/1.53/13  & 0.35/0.96/12  & 0.12/1.62/18  & 0.26/1.17/13  \\ \cline{2-7} 
                     & T06 vs T20+PVP & \textbf{0.15/1.99/1}  & 0.05/-2.18/11 & 0.16/-1.52/9  & 0.16/-1.45/17 & 0.22/-1.27/13 \\ \hline
\multirow{3}{*}{$PL_{ps}$} & T06 vs T20     & 0.52/0.66/8   & \textbf{0.21/1.57/1}  & 0.80/-0.25/7  & 0.36/0.96/8   & 0.16/-1.49/12 \\ \cline{2-7} 
                     & T20 vs T20+PVP & 0.21/1.30/10  & \textbf{0.90/0.01/1}  & 0.51/-0.66/10 & 0.39/0.89/12  & 0.71/-0.36/13 \\ \cline{2-7} 
                     & T06 vs T20+PVP & 0.48/0.72/10  & \textbf{0.26/1.27/1}  & 0.60/-0.53/9  & 0.79/-0.26/8  & 0.56/0.59/13  \\ \hline
\multirow{3}{*}{$PL_{rs}$} & T06 vs T20     & 0.23/1.28/8   & \textbf{0.42/0.64/1}  & 0.79/0.26/7   & 0.37/-0.94/8  & 0.36/0.94/12  \\ \cline{2-7} 
                     & T20 vs T20+PVP & 0.75/-0.32/10 & \textbf{0.02/5.10/1}  & 0.71/0.37/10  & 0.59/0.53/12  & 0.17/1.43/13  \\ \cline{2-7} 
                     & T06 vs T20+PVP & 0.25/-1.19/10 & \textbf{0.41/0.67/1}  & 0.98/0.01/9   & 0.24/1.25/8   & 0.46/0.75/13  \\ \hline
\multirow{3}{*}{$Re_{p}$} & T06 vs T20     & \textbf{$<$0.001/10.12/1}   & $<$0.001/5.04/10  & $<$0.001/4.99/9   & \textbf{$<$0.001/12/1}  & $<$0.001/10.21/12  \\ \cline{2-7} 
                     & T20 vs T20+PVP & \textbf{$<$0.001/10.68/1} & 0.005/3.29/13  & 0.003/3.59/12  & \textbf{$<$0.001/10.92/1}  & $<$0.001/6.05/13  \\ \cline{2-7} 
                     & T06 vs T20+PVP & \textbf{$<$0.001/10.6/1} & 0.07/-1.95/11  & 0.03/-2.50/9   & \textbf{$<$0.001/10.37/1}   & $<$0.001/-4.37/13  \\ \hline
\multirow{3}{*}{$Re_{r}$} & T06 vs T20     & 0.24/-1.20/13   & 0.58/0.56/10  & 0.81/-0.23/9   & 0.47/0.72/15  & 0.85/-0.15/12  \\ \cline{2-7} 
                     & T20 vs T20+PVP & 0.49/-0.69/17 & 0.467/-0.73/13  & 0.20/-1.34/12  & 0.35/-0.95/18  & 0.54/-0.61/13  \\ \cline{2-7} 
                     & T06 vs T20+PVP & 0.44/0.78/14 & 0.27/-1.14/11  & 0.32/-1.03/9   & 0.08/-1.83/17   & 0.54/-0.62/13  \\ \hline
\multirow{3}{*}{$St$} & T06 vs T20     & 0.38/0.89/13   & 0.50/-0.68/10  & 0.61/0.52/9   & 0.48/-0.71/15  & 0.75/0.32/12  \\ \cline{2-7} 
                     & T20 vs T20+PVP & 0.81/0.24/17 & 0.96/-0.04/13  & 0.28/1.12/12  & 0.62/0.49/18  & 0.75/0.32/13  \\ \cline{2-7} 
                     & T06 vs T20+PVP & 0.39/-0.87/14 & 0.51/0.67/11  & 0.59/0.55/9   & 0.15/1.49/17   & 0.99/0.01/13  \\ \hline
   
\end{tabular}
}
\caption{\label{tab: Stats} Statistical comparisons between treatments for each pleopod. For the t-test, the reported values are the \textit{p}-value (\textit{p}), t-value (t), and degrees of freedom (\textit{df}). When a Kruskal-Wallis test was performed (\textbf{bold}), the reported values were \textit{p}-value, $\chi^2$, and degrees of freedom.}
\end{table}

\subsection{Pleopods kinematics}

Low temperatures are known to reduce muscle contraction rate, resulting in low frequencies and swimming velocities \cite{koumoundouros_effect_2002}. However, the advance ratio suggests that neither slow appendage beating nor exposure to high viscosity fluid impacts swimming performance. In this section, we explore all the kinematic variables describing the pleopod motion to investigate how shrimp reacts to environmental changes from a propulsion standpoint alone. Starting with the beat frequency (Figure \ref{fig:KineVar}A), one-way ANOVAs show that there are no significant differences between pleopods (T06-ANOVA, $p = 0.87, F = 0.31, df = 4$ / T20-ANOVA, $p = 0.61, F = 0.68, df = 4$ / T20+PVP-ANOVA, $p = 0.93, F = 0.21, df = 4$). This is the only pleopod variable that shows no change between neighboring paddles. When comparing $f$ between treatments, it is clear that low temperature causes a decrease in beat frequency (see Table \ref{tab: Stats} $f$, T06 \textit{vs} T20 and T06 \textit{vs} T20+PVP). Looking at the same temperature cases (T20 \textit{vs} T20+PVP), there is no significant difference; beat frequency changes are only due to physiological effects (temperature). 

The protopodite motion amplitude $\alpha$ is significantly different between pleopods (T06-ANOVA, $p = <0.001, F = 8.38, df = 4$ / T20-ANOVA, $p = <0.001, F = 13.86, df = 4$ / T20+PVP-ANOVA, $p = <0.001, F = 14.87, df = 4$), P1 having the smaller rage of motion $(\sim 60^\circ)$ to P5 having the largest $(\sim 90^\circ)$ (Figure \ref{fig:KineVar}B). When comparing treatments, there is a decreasing trend from T06, T20, to T20+PVP, but only the high-viscosity cases T06 and T20+PVP are significantly different (see Table 1; P1 and P4 have a $p < 0.05$). This suggests that the change in range of motion ($\alpha$) is triggered by temperature, not viscosity. Shrimp actively maintain $\alpha$ when exposed to low temperatures in response to the lower beat frequency; however, an artificial high viscosity at $20 ^\circ $ C with no change in beat frequency (Figure \ref{fig:KineVar}A), decreases the pleopod motion (i.e., lower $\alpha$). Nonetheless, there is no significant difference between T06 and T20; thus, shrimp do not modify $\alpha$ when exposed to a change in temperature. 

The amplitude $\beta$ is a variable that traditionally assumes the distal appendage is rigid when, in fact, it is distinctly flexible (Figure \ref{fig:KineVar}E). The beat cycle is a highly dynamic process, during which the point of maximum curvature travels through the pleopod (Figure \ref{fig:KineVar}F). For this reason, $\beta$ is not an accurate description of the distal appendage kinematics; it is just a rough estimate of the range of motion. One-way ANOVAs show significant differences between pleopods (T06-ANOVA, $p = <0.001, F = 10.86, df = 4$ / T20-ANOVA, $p = <0.001, F = 8.62, df = 4$ / T20+PVP-ANOVA, $p = 0.01, F = 13.04, df = 4$). Just as for $\alpha$, there is an increasing trend from P1 $(\sim 65^\circ)$ to P5 $(\sim 90^\circ)$. Figure \ref{fig:KineVar}D shows a decrease in $\beta$, with T20 having the highest value, followed by T20+PVP, and T06 having the lowest. This agrees with the argument that $\beta$ is a passive result of fluid-structure interactions during forward swimming \cite{tack_going_2024}. T20 is the low viscosity case (i.e., less motion resistant) and a high beating frequency (i.e., fast motion), which results in the highest range of motion. T20+PVP has high viscosity (i.e., more motion resistant) and high beat frequency; the added resistance results in a lower value of $\beta$. Finally, T06 has high viscosity (i.e., more resistant) at low beating frequency (i.e., slow motion), resulting in the lowest $\beta$. Although T06 \textit{vs.} T20 are statistically different (see Table \ref{tab: Stats}), $\beta$ is not does not accurately describe the motion of the highly flexible biramous distal appendage and is likely not actively controlled by the shrimp.

\begin{figure}[hbt!]
\centering
\includegraphics[width=\linewidth]{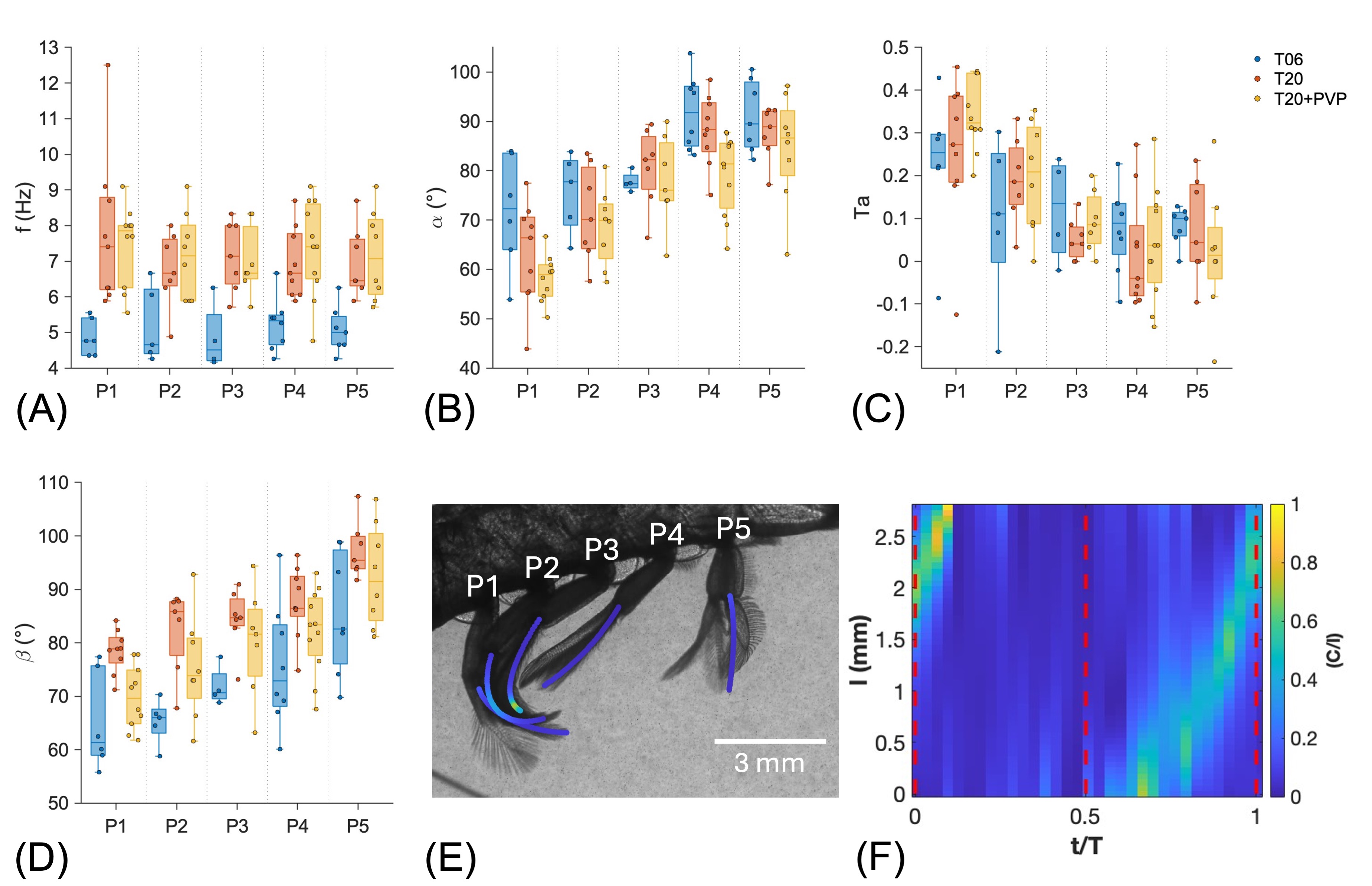}
\caption{Pleopod kinematic variables. (A) Beat frequency $f$. (B) Stroke amplitude $\alpha$. (C) Temporal asymmetry $Ta$. (D) distal appendage amplitude $\beta$. (E) Snapshot of the pleopods during forward swimming, displaying the normalized curvature on top of each distal appendage. (F) P3 normalized curvature values during an entire beat cycle; red dotted lines indicate the power and recovery stroke transition. For each pleopod variable, each observation corresponds to a complete beat cycle. The number of observations for \textcolor{blue}{T06}: $n_{P1} = 7$, $n_{P2} = 7$, $n_{P3} = 7$, $n_{P4} = 8$, and $n_{P5} = 8$, for \textcolor{red}{T20}: $n_{P1} = 9$, $n_{P2} = 8$, $n_{P3} = 8$, $n_{P4} = 10$, and $n_{P5} = 9$, and for \textcolor{Goldenrod}{T20+PVP}: $n_{P1} = 10$, $n_{P2} = 8$, $n_{P3} = 7$, $n_{P4} = 11$, and $n_{P5} = 9$.  The boxplots display the median, the lower and upper quartiles, outliers, and the minimum and maximum values that are not outliers.}
\label{fig:KineVar}
\end{figure}

The temporal asymmetry is the last kinematic parameter we considered for an individual pleopod (Figure \ref{fig:KineVar}C). This parameter is zero for a time-symmetric beat cycle and one for an infinitely fast power stroke. Table \ref{tab: Stats} shows no statistical difference between treatments for the $Ta$ parameter. From one-way ANOVAs, we found significant differences between pleopods for the T20 and T20+PVP cases (T20-ANOVA, $p = <0.001, F = 5.38, df = 4$ / T20+PVP-ANOVA, $p = <0.001, F = 12.25, df = 4$), and no significant difference for the T06 case (T06-ANOVA, $p = 0.26, F = 1.39, df = 4$). Despite not having a significant difference in the T06 case, it is clear from Figure \ref{fig:KineVar}C that all treatments show a decreasing trend in $Ta$ from P1 $(\sim 0.35)$ to P5 $(\sim 0.05)$. Simply put, the anterior pleopods have faster power strokes than the posterior. 

Shrimp-like organisms are known to display complete or hybrid metachronal locomotion, in which the organisms complete metachronal power strokes, but quasi-synchronous recovery strokes  \cite{hanson_mantis_2023}. Through the use of a simplified robotic platform, Ford et al.\cite{ford_hybrid_2021} showed that hybrid metachronal rowing allowed to operate the pleopods at larger stroke amplitudes ($\alpha$), resulting in higher swimming speeds. However, the control strategy behind the hybrid coordination is unclear --- varying the $Ta$ between consecutive pleopods (see Figure \ref{fig:KineVar}C) could be the driving mechanism of the coordination.
When the phase lag is measured at the beginning of the power stroke (Figure \ref{fig:PL}A), there is no significant difference between neighboring pleopods (T06-ANOVA, $p = 0.15, F = 6.71, df = 4$ / T20-ANOVA, $p = 0.23, F = 1.47, df = 4$ / T20+PVP-ANOVA, $p = 0.22, F = 3.30, df = 4$) resulting in metachronal coordination during the power stroke. On the other hand, if we take the beginning of the recovery stroke as the time reference (Figure \ref{fig:PL}B), there is a significant difference in the phase lag (T06-ANOVA, $p = 0.04, F = 9.12, df = 4$ / T20-ANOVA, $p = <0.001, F = 18.11, df = 4$ / T20+PVP-ANOVA, $p = <0.001, F = 19.76, df = 4$). The phase lag between P1 and P5 shows the increased value characteristic of the hybrid metachronal motion. P1 and P2  have faster power strokes and slower recovery strokes than the rest of the pleopods (Figure \ref{fig:KineVar}C). This combination results in the anterior pleopods (P1 and P2) finishing their power strokes fast enough to catch up with the recovery of the rest of the pleopods, resulting in a synchronous-looking recovery. 

\begin{figure}[hbt!]
\centering
\includegraphics[width=\linewidth]{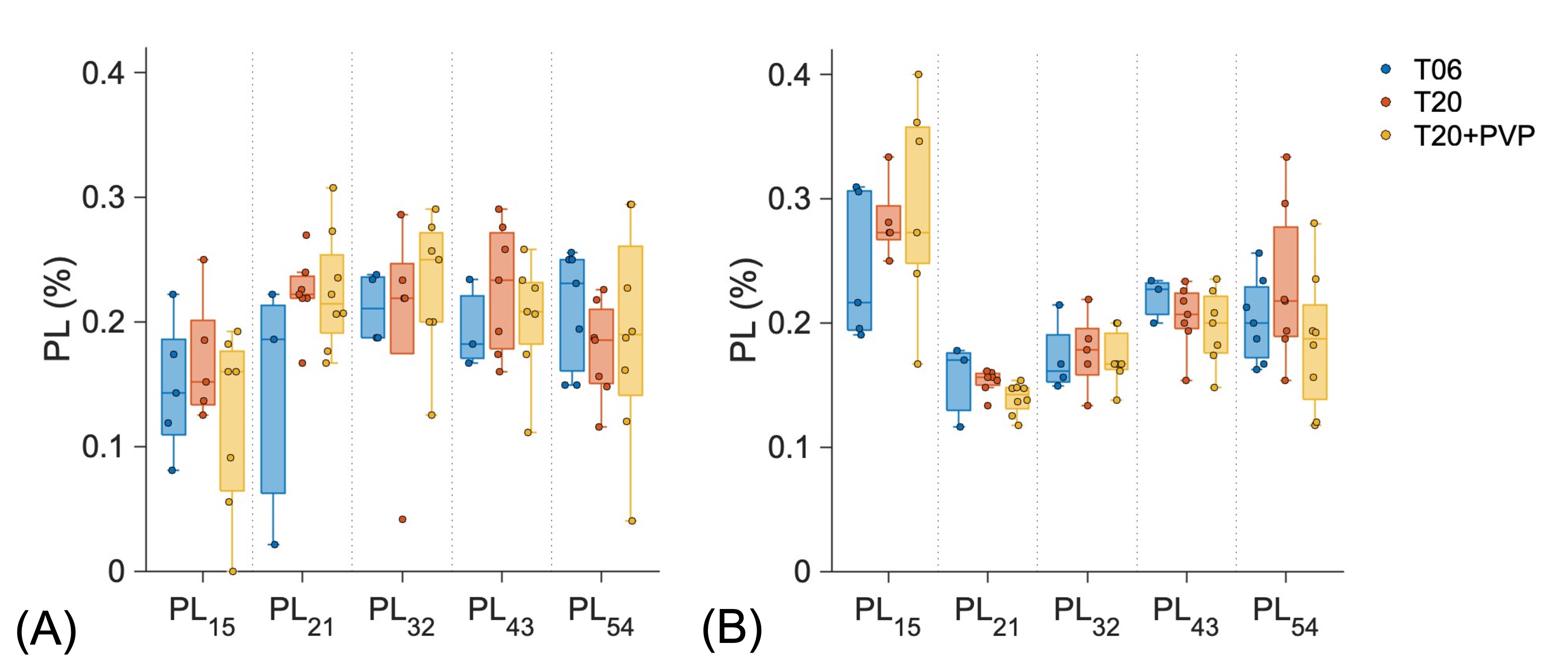}
\caption{Phase lag (PL) between adjacent pleopod strokes (reported as a fraction of a complete cycle). (A) Phase lag using the beginning of the power stroke as a reference time instant. (B) Phase lag value considering the beginning of the recovery stroke as the reference time instant. The number of observations for phase lag calculations can be smaller than the rest of the kinematic variables due to the additional restriction of consecutive beat cycles between neighboring pleopods. The number of observations for \textcolor{blue}{T06}: $n_{P1} = 7$, $n_{P2} = 7$, $n_{P3} = 7$, $n_{P4} = 7$, and $n_{P5} = 8$, for \textcolor{red}{T20}: $n_{P1} = 7$, $n_{P2} = 8$, $n_{P3} = 8$, $n_{P4} = 8$, and $n_{P5} = 9$, and for \textcolor{Goldenrod}{T20+PVP}: $n_{P1} = 9$, $n_{P2} = 8$, $n_{P3} = 7$, $n_{P4} = 7$, and $n_{P5} = 9$. The boxplots display the median, the lower and upper quartiles, outliers, and the minimum and maximum values that are not outliers.}
\label{fig:PL}
\end{figure}

\subsection{Flow features}

The Reynolds number ($Re$) and the Strouhal number $(St)$ are canonical dimensionless quantities used to describe the fluid dynamics of swimming. When the primary source of propulsion is the motion of appendages, we need to consider two Reynolds numbers: the body-based $(Re_b = V_bL_b/\nu)$ and the appendage-based ($Re_p = 2 \pi f l^2 /\nu$, where $l = L_p+L_e$). Figure \ref{fig:swiming}A shows the body Reynolds number for our three treatments. As expected, being a function of body speed (Figure \ref{fig: BodyKine}A) and the change in viscosity, the $Re_b$ of all treatments are statistically different between each other (T06 \textit{vs.} T20 t-test, $p = <0.001, t = 5.99, df = 12$ / T06 \textit{vs.} T20+PVP t-test, $p = 0.005, t = -3.36, df = 12$ / T20 \textit{vs.} T20+PVP t-test, $p = 0.009, t = 3.07, df = 12$). This trend is also clear when evaluating the appendage Reynolds number (Figure \ref{fig:swiming}B). Despite these differences on $Re_b$ and $Re_p$, their ratio (Figure \ref{fig:swiming}C) does not show differences between treatments (Table \ref{tab: Stats}). Like the Reynolds number ratio, the Strouhal number ($St = fl/V_b$, Figure \ref{fig:swiming}D) does not show a difference between treatments (Table \ref{tab: Stats}). This shows that despite the slower movements caused by the temperature decrease, the shrimp always swim at a fixed advance ratio---we will use this result to simulate steady-state swimming conditions and disentangle the physical and physiological effects in swimming performance.

\begin{figure}[htb!]
\centering
\includegraphics[width=\linewidth]{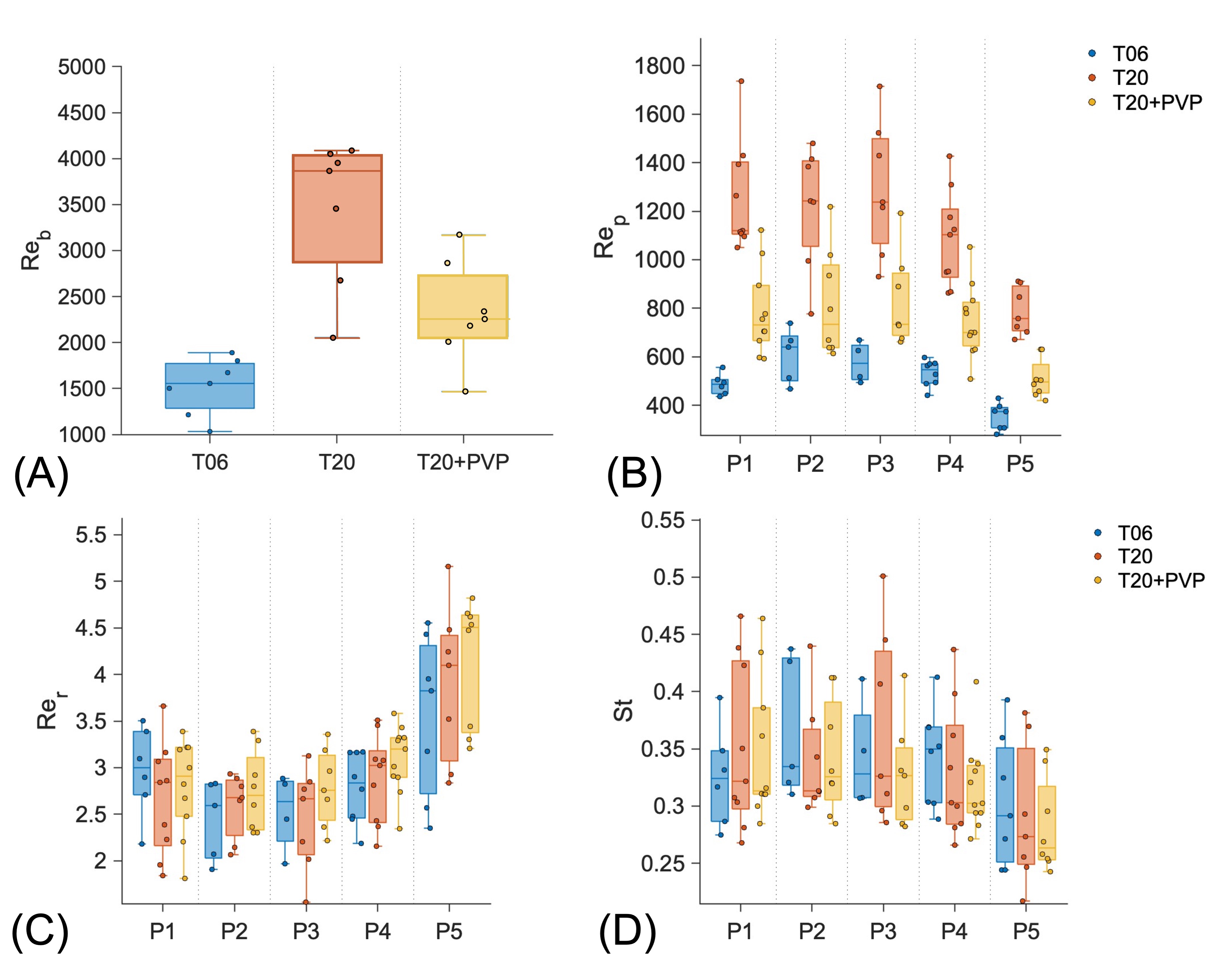}
\caption{Shrimp characteristic swimming numbers.(A) Body-based Reynolds number ($Re_b$) for the three treatments. (B) Pleopod-based Reynolds number ($Re_p$) for each pleopod. (C) The ratio between body and pleopod-based Reynolds numbers ($Re_r = Re_b/Re_p$). (C) Shrimp Strouhal number for each pleopod. The boxplots displays the median, the lower and upper quartiles, outliers, and the minimum and maximum values that are not outliers.}
\label{fig:swiming}
\end{figure}

 We used a three-dimensional CFD model of a swimming shrimp to investigate the fluid dynamics behind the gait resiliency to environmental changes. While this model does not explicitly consider the physiological effects (i.e., muscle function), it allows for a parametric study of forward-swimming organisms. The model was informed with shrimp kinematics from the T20 treatment, such that $\alpha$, $Ta$, $\beta$, and $PL$ are held constant while we varied the beat frequency and the water viscosity (i.e., the parameters affected by water temperature). During the simulation, the shrimp model is fixed at the center of a flow domain, and an inflow velocity is given to the front boundary of the simulation domain. To simulate steady-state swimming, we set the Strouhal number to match the mean experimental value $(St = 0.33)$, which effectively fixes the advance ratio $(J = 1/2 \alpha St)$. Therefore, we can calculate the inflow velocity for different beat frequencies as $V_b = f l / St$. We assume that shrimp reach this Strouhal number value during steady-state swimming because $St$, $Re_{r}$, and $J$ show no statistical difference between treatments (Figure \ref{fig: BodyKine}B and \ref{fig:swiming}C-D). 

 Under natural conditions, reducing the temperature decreases the beat frequency and increases the fluid viscosity--- this combined effect results in a drastic reduction of the vortex signature in the near field to the pleopods (Figure \ref{fig:Vortex}A and C). The effect is also evident when plotting the pressure field (Figure \ref{fig:Vortex}B and D), where the pressure signature is lower when we have a case representing both the physiological and physical effects. On the other hand, when we isolate viscosity effects (Figure \ref{fig:Vortex}A-B and E-F), the vortex structures and the pressure fields produced by the pleopods are less impacted. To interpret these results in terms of propulsion performance, we calculated the average thrust-to-power ratio (i.e., thrust efficiency variable) for the expected range of $Re_p = 2 \pi f l^2 /\nu$ of the organisms considered in this work (Figure \ref{fig:Eff}A). The plot shows how decreasing the Reynolds number results in a lower thrust performance; therefore, lowering the beat frequency (i.e., the physiological effect of cold water) results in a performance loss. The color stars in Figure \ref{fig:Eff} highlight the average $Re_p$ for the experimental conditions. We can observe going from T20 (red) to T06 (blue) (i.e., coupled physiological and physical effects) results in a performance loss ($47.77 \%$). However, when isolating the physical effects of viscosity (T20 to T20+PVP), the thrust-to-power ratio only varies by $8.26 \%$. On the other hand, lift production (Figure \ref{fig:Eff}B) comparing T06 \textit{vs.} T20, there is a performance loss of $27.28 \%$ and only a $4.35 \%$ loss for T20 \textit{vs.} T20+PVP. The environmental effects impact lift production less aggressively than thrust generation. These results align with the propulsion efficiency evaluation by \textit{Herrera et al.}. \cite{herrera-amaya_propulsive_2024} for rowing at intermediate Reynolds numbers. In this case, the appendage shape is the primary factor in generating lift, while changes in speed and viscosity come in second. Shrimp are negatively buoyant \cite{ford_role_2021}; their ability to generate lift under any environmental condition is crucial for maintaining a constant vertical position.
 
\begin{figure}[htb!]
\centering
\includegraphics[width=30pc]{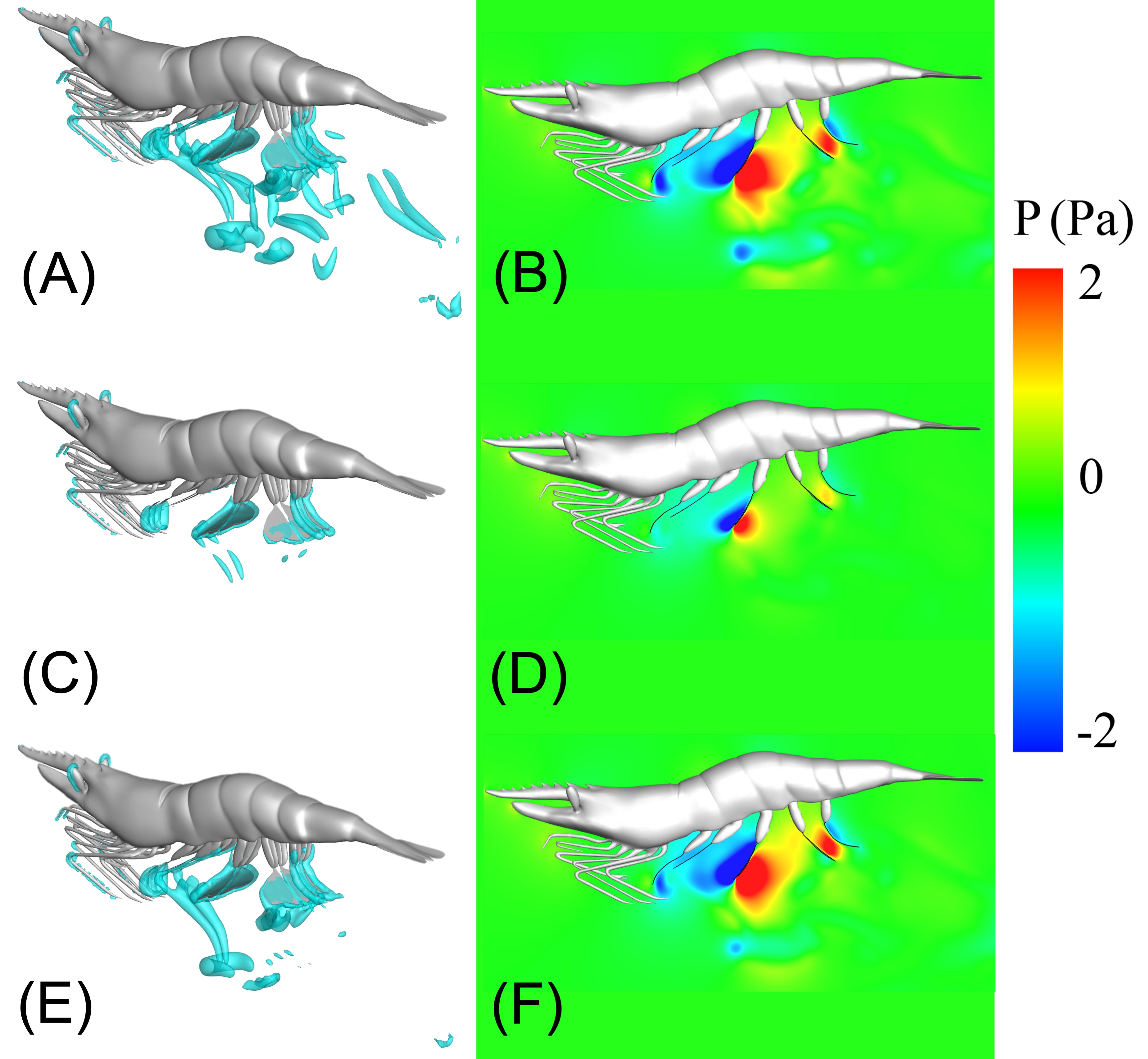}
\caption{Vortex structures displayed based on a non-dimensional Q-criterion ($QV_b^2/l^2=40$) and pressure field view at the domain center for the time instant $(t/T=0.8)$. (A-B) Case representing \textcolor{red}{T20}: viscosity $\nu = 1.017 \ mm^2/s$, and $f = 9 Hz$. (C-D) Case representing \textcolor{blue}{T06}: viscosity $\nu = 1.519 \ mm^2/s$, and $f = 6 Hz$. (E-F) Case representing \textcolor{Goldenrod}{T20+PVP}: viscosity $\nu = 1.519 \ mm^2/s$, and $f = 6 Hz$.} 
\label{fig:Vortex}
\end{figure}

\begin{figure}[htb!]
\centering
\includegraphics[width=\linewidth]{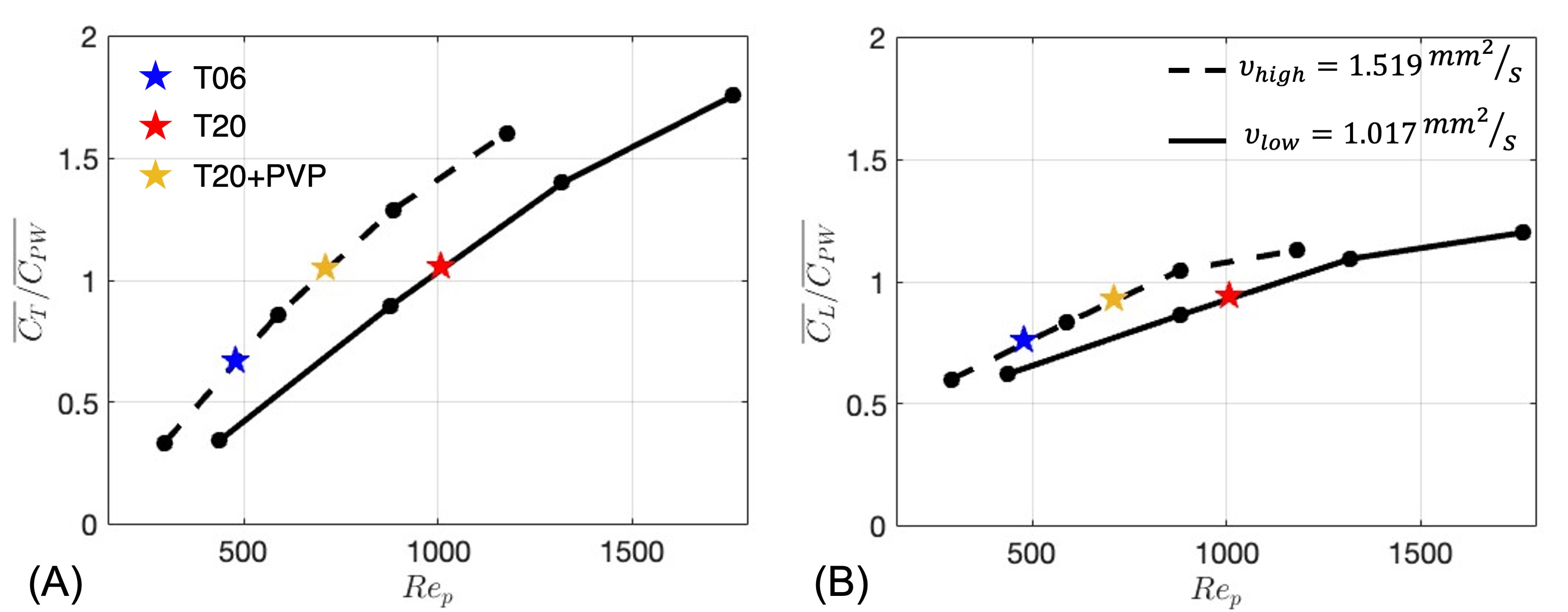}
\caption{Steady-state swimming performance of simulated grass shrimp. (A) Cycle average thrust-to-power ratio. (B) Cycle average lift-to-power ratio. For ocean water viscosities at $20 ^\circ C$ $(\nu_{low})$ and  at $6 ^\circ C$ $(\nu_{high})$.}
\label{fig:Eff}
\end{figure}

\newpage
\section{Discussion}

Our results show that metachronal locomotion in shrimp-like organisms has evolved to be highly resilient to environmental water-dependent properties. Experimental observations of marsh grass shrimp revealed that the forward swimming gait does not change significantly when exposed to natural temperature extremes $(6^\circ - 20^\circ C)$, while maintaining similar swimming performance (i.e., advance ratio) despite the forced decrease in beat frequency when exposed to cold water (physiological effects).

Among the studied kinematic parameters characterizing forward swimming, the only variable showing a significant change between treatments is the amplitude $\beta$. While this parameter has been traditionally used to estimate the motion of the distal pleopod, our results show that it does not accurately describe the motion of the distal appendage. Rather, it is an estimate of the range of motion assuming the distal appendage is rigid--- Figure \ref{fig:KineVar}E-F shows the complex bending behavior during the beat cycle. Nonetheless, the reduction of $\beta$ angle with an increase of water viscosity (Figure \ref{fig:KineVar}D) agrees with the hypothesis that the asymmetric beat cycle is a passive result of the geometry and material properties of the distal appendage interacting with the surrounding fluid \cite{tack_going_2024}. Therefore, the distal bending of the appendage is not considered to be actively controlled by the animals.

The connection between appendage morphology and kinematics is not well understood; pleopods have different lengths (Figure \ref{fig:Morphology}D-E) and kinematics (Figure \ref{fig:KineVar}B-D). The second pleopod is the largest, $\alpha$ and $\beta$ increase from P1 to P5, and $Ta$ decreases from P1 to P5. This is the first time temporal asymmetry has been evaluated with a parameter directly relating power and recovery stroke times $(Ta = t_r-t_p/t_r+t_p)$, in shrimp. Figure \ref{fig:KineVar}A shows that all pleopods have the same beat frequency (i.e., same stroke time), but the temporal asymmetry (Figure \ref{fig:KineVar}C) aggressively changes from faster power strokes (P1) to almost symmetric strokes (P5). When examining the phase lag (Figure \ref{fig:PL}), we realized that the decreasing value of $Ta$ could be the driving mechanism for the hybrid metachronal swimming. Posterior pleopods start the metachronal wave with longer power strokes, and the anterior pleopods finish with shorter power strokes. As a result, the anterior pleopods catch up on the recovery strokes of the rest of the pleopods. Simply put, the power strokes look metachronal, and the recovery seems synchronous. This hybrid coordination has been shown to increase the swimming speed of shrimp \cite{ford_hybrid_2021, hanson_mantis_2023}. Going forward, we aim to develop a shrimp-inspired robotic platform to disentangle the combined effects of morphology and kinematic differences between consecutive pleopods in propulsion performance. 

Metachronal coordination is known for its high efficiency at intermediate \emph{Re} \cite{barlow_water_1993}---however, its adaptability to shifting ocean fluid properties remained an open question. Metachronal crustaceans are ecologically diverse (large number and kind of habitats), taxonomically diverse (many genera and species), and biologically productive (sizable biomass) \cite{bauer_shrimps_2023}. This is evidence of their ability to adapt to vastly different water temperatures. This study employed a lab-based CFD model to decouple the physiological and physical effects of water temperature on shrimp swimming performance. The results show that the shrimp metachronal gait is robust to viscosity changes (Figure \ref{fig:Eff}), resulting in a locomotor strategy that is well-suited to adapt to water viscosity fluctuations, both from abiotic factors such as temperature or salinity changes and biotic like phytoplankton blooms \cite{jenkinson_oceanographic_1986,seuront_biologically_2006,seuront_increased_2008}. Our results highlight the potential of engineering crustacean-inspired underwater vehicles. Robotic devices are much less affected by temperature changes (i.e., no physiological effects). In addition, as confirmed by our work, metachronal underwater drones have the potential to navigate environments with variable viscosity without modifications to their control algorithm, simplifying the task of navigating during harsh environmental conditions--both natural, like phytoplankton blooms or human-produced, like oil spills.

\section{Acknowledgements}
We gratefully acknowledge Sara O. Santos for her insightful discussions that strengthened our work. We also thank Marjorie Bradley for her assistance with coordination and administrative support throughout this research. Additionally, the authors would like to acknowledge the funding support from the NASA Ocean Biology and Biogeochemistry Program (80NSSC22K0284) to M. M. Wilhelmus, the National Science Foundation (CBET-2451990) to C. Li, and the Presidential Postdoctoral Fellowship at Brown University to A. Herrera-Amaya.

\section{Author contributions}
A.H.A. conceptualization, formal analysis, methodology, funding acquisition, writing --- original draft, review, and editing; N.B.T. conceptualization, experiments, methodology, writing --- review and editing; Z.L. numerical simulations, formal analysis, writing --- review and editing; C.L. conceptualization, formal analysis, methodology, supervision, funding acquisition, writing --- review and editing; M.M.W. conceptualization, formal analysis, methodology, supervision, funding acquisition, resources, project administration, writing--- review and editing.\\

The authors declare no competing financial interest.\\

%\textit{Data availability}— The data that support the findings of this study are available at the following URL/DOI: https://drive.google.com/drive/folders/1Es9rLIkfcBqm3OA9-b7x32o49ux7LWec?usp=sharing. Data will be posted in an open repository upon acceptance of the manuscript.

\newpage

\section{Appendixes}
Table \ref{tab:Test1} shows the normality and variance tests for the body variables; we recorded seven animals per treatment (n). Tables \ref{tab:Test3} and \ref{tab:Test2} show the assumption tests for the appendage variables; here, each observation corresponds to a complete recorded beat cycle of the appendage. The p-values of data sets violating assumptions are highlighted.

\begin{table}[hbt!]
\centering

\begin{tabular}{|c|c|cl|cc|}
\hline
\multirow{2}{*}{Treatment}    & \multirow{2}{*}{n}     & \multicolumn{2}{c|}{Normality}                            & \multicolumn{2}{c|}{Variances}                                         \\ \cline{3-6} 
                              &                        & \multicolumn{1}{c|}{$V_b$}  & \multicolumn{1}{c|}{$R_b$}  & \multicolumn{1}{c|}{$V_b$}                   & $R_b$                   \\ \hline
T06                           & 7                      & \multicolumn{1}{c|}{0.9849} & \multicolumn{1}{c|}{0.6982} & \multicolumn{1}{c|}{\multirow{3}{*}{0.4710}} & \multirow{3}{*}{0.1206} \\ \cline{1-4}
T20                           & \multicolumn{1}{l|}{7} & \multicolumn{1}{l|}{0.0506} & 0.0628                      & \multicolumn{1}{c|}{}                        &                         \\ \cline{1-4}
\multicolumn{1}{|l|}{T20+PVP} & \multicolumn{1}{l|}{7} & \multicolumn{1}{l|}{0.9849} & 0.8442                      & \multicolumn{1}{c|}{}                        &                         \\ \hline
\end{tabular}

\caption{\label{tab:Test1} Shapiro-Wilk normality test and Levene's variances test \textit{p}-values for swimming body variables.}
\end{table}

\begin{table}[hbt!]
\begin{tabular}{|l|l|l|l|l|l|l|l|l|l|l|l|}
\hline
Treatment & Pleopod & J      & f               & $\alpha$        & Ta              & $PL_{rs}$       & $PL_{ps}$       & $\beta$         & $Re_p$          & $Re_r$ & St     \\ \hline
          & P1      & 0.4959 & 0.0682          & \textbf{0.0373} & 0.4279          & 0.0707          & 0.3386          & \textbf{0.0157} & \textbf{0.0394} & 0.3365 & 0.2393 \\ \cline{2-12} 
          & P2      & 0.9610 & 0.7914          & 0.6421          & 0.3998          & \textbf{0.0065} & \textbf{0.0188} & 0.3013          & 0.2960          & 0.7066 & 0.5753 \\ \cline{2-12} 
T06, T20, & P3      & 0.6710 & 0.9285          & 0.1585          & \textbf{0.0280} & 0.8227          & 0.3524          & 0.2528          & 0.1101          & 0.4631 & 0.0642 \\ \cline{2-12} 
T20+PVP   & P4      & 0.8133 & 0.2300          & 0.7393          & 0.7035          & 0.5905          & 0.5879          & 0.2452          & \textbf{0.0272} & 0.5246 & 0.2543 \\ \cline{2-12} 
          & P5      & 0.3344 & \textbf{0.0453} & 0.3014          & 0.1658          & 0.3994          & 0.2495          & 0.0570          & 0.1752          & 0.7511 & 0.5130 \\ \hline
T06       &         & 0.7660 & 0.2945          & 0.0520          & 0.2689          & \textbf{0.0255} & \textbf{0.0460} & \textbf{0.0417} & \textbf{0.0137} & 0.4631 & 0.4990 \\ \cline{1-1} \cline{3-12} 
T20       & P1-P5   & 0.1128 & 0.2059          & 0.2768          & 0.2283          & \textbf{0.0198} & 0.1993          & 0.4425          & 0.2730          & 0.7511 & 0.3226 \\ \cline{1-1} \cline{3-12} 
T20+PVP   &         & 0.0524 & 0.8298          & 0.3339          & 0.4581          & \textbf{0.0028} & 0.4057          & 0.7428          & 0.1079          & 0.0844 & 0.3290 \\ \hline
\end{tabular}
\caption{\label{tab:Test3} Levene's variances test \textit{p}-values for appendage variables. First is the pleopod variances between treatments (T06, T20, T20+PVP). Then, the variances between each treatment's five pleopods (P1-P5). Bold numbers highlight the data sets that fail the variance test.}
\end{table}

\begin{table}[hbt!]
\begin{tabular}{|l|c|c|c|c|c|c|c|c|c|c|c|}
\hline
Treatment & Pleopod & J      & f               & $\alpha$ & Ta     & $PL_{rs}$ & $PL_{ps}$ & $\beta$         & $Re_p$          & $Re_r$ & St     \\ \hline
          & P1      & 0.8466 & 0.2244          & 0.6345   & 0.1265 & 0.0725    & 0.9928    & 0.1352          & 0.8192          & 0.7864 & 0.8738 \\ \cline{2-12} 
          & P2      & 0.5855 & 0.2044          & 0.6620   & 0.5809 & 0.2159    & 0.3241    & 0.7299          & 0.6595          & 0.1616 & 0.1589 \\ \cline{2-12} 
T06       & P3      & 0.5136 & 0.1536          & 0.4777   & 0.4897 & 0.1903    & 0.0513    & 0.2016          & 0.4376          & 0.4321 & 0.2390 \\ \cline{2-12} 
          & P4      & 0.4920 & 0.5343          & 0.3396   & 0.8988 & 0.3609    & 0.4125    & 0.8485          & 0.5183          & 0.2052 & 0.4411 \\ \cline{2-12} 
          & P5      & 0.4013 & 0.7003          & 0.4416   & 0.3751 & 0.7720    & 0.0680    & 0.3956          & 0.4965          & 0.7791 & 0.6148 \\ \hline
          & P1      & 0.3384 & \textbf{0.0432} & 0.7473   & 0.1092 & 0.2326    & 0.3498    & 0.8209          & \textbf{0.0430} & 0.7916 & 0.2066 \\ \cline{2-12} 
          & P2      & 0.3464 & 0.6507          & 0.6792   & 0.9960 & 0.0565    & 0.1561    & \textbf{0.0253} & 0.3478          & 0.1130 & 0.0690 \\ \cline{2-12} 
T20       & P3      & 0.3558 & 0.5383          & 0.5339   & 0.5430 & 0.9683    & 0.0847    & 0.1491          & 0.8980          & 0.7065 & 0.2417 \\ \cline{2-12} 
          & P4      & 0.2662 & 0.2912          & 0.9706   & 0.0510 & 0.3342    & 0.5177    & 0.7391          & 0.4688          & 0.5151 & 0.2770 \\ \cline{2-12} 
          & P5      & 0.7183 & 0.4070          & 0.1544   & 0.6252 & 0.3726    & 0.8135    & 0.3483          & 0.2942          & 0.7214 & 0.3301 \\ \hline
          & P1      & 0.0727 & 0.1912          & 0.9923   & 0.3489 & 0.8371    & 0.2750    & 0.3980          & 0.2077          & 0.3186 & 0.0520 \\ \cline{2-12} 
          & P2      & 0.8129 & 0.2941          & 0.8507   & 0.5694 & 0.3801    & 0.6732    & 0.8899          & 0.1373          & 0.2241 & 0.2103 \\ \cline{2-12} 
T20+PVP   & P3      & 0.6300 & 0.1844          & 0.7739   & 0.8782 & 0.1800    & 0.4177    & 0.9300          & 0.1709          & 0.9734 & 0.2810 \\ \cline{2-12} 
          & P4      & 0.8676 & 0.6031          & 0.1066   & 0.8969 & 0.9343    & 0.4259    & 0.7054          & 0.8867          & 0.2549 & 0.1537 \\ \cline{2-12} 
          & P5      & 0.9079 & 0.3975          & 0.5101   & 0.7419 & 0.6640    & 0.7014    & 0.6290          & 0.1174          & 0.0579 & 0.0659 \\ \hline
\end{tabular}
\caption{\label{tab:Test2} Shapiro-Wilk normality test \textit{p}-values for appendage variables. Bold numbers highlight the data sets that fail the normality test.  }
\end{table}

\newpage

\bibliography{Refe}% Produces the bibliography via BibTeX.

\end{document}